\documentclass[aps,prb,preprint,superscriptaddress]{revtex4-2}
\usepackage{CJK}
\usepackage{natbib}
\usepackage{graphicx}
\usepackage{amsmath}
\usepackage{amssymb}
\usepackage {color} 
\usepackage{kotex}
\UseRawInputEncoding
\begin{document}
\begin{CJK*}{GB}{} 


\title{Plasmon thermal conductivity of thin Au and Ag films}


\author{Dong-min Kim}
\affiliation{Department of Mechanical Engineering, Korea Advanced Institute of Science and Technology, Daejeon 34141, South Korea}
\affiliation{Center for Extreme Thermal Physics and Manufacturing, Korea Advanced Institute of Science and Technology, Daejeon 34141, South Korea}

\author{Jeongmin Nam}
\affiliation{Department of Mechanical Engineering, Korea Advanced Institute of Science and Technology, Daejeon 34141, South Korea}
\affiliation{Center for Extreme Thermal Physics and Manufacturing, Korea Advanced Institute of Science and Technology, Daejeon 34141, South Korea}

\author{Bong Jae Lee}
\email{bongjae.lee@kaist.ac.kr}
\affiliation{Department of Mechanical Engineering, Korea Advanced Institute of Science and Technology, Daejeon 34141, South Korea}
\affiliation{Center for Extreme Thermal Physics and Manufacturing, Korea Advanced Institute of Science and Technology, Daejeon 34141, South Korea}



\maketitle
\end{CJK*}
\date{\today}

\begin{abstract}
We investigated the thermal conductivity of surface plasmon polaritons (SPPs) propagating along thin Au and Ag films on a SiO$_2$ substrate with a Ti adhesive layer. To determine the propagation length and skin depth of SPPs along Au and Ag thin films, we numerically solved the dispersion relation while considering the size effect of the permittivity of metal. Additionally, we derived the spatial distribution of SPPs along the film thickness to analyze the effect of the Ti adhesive layer on the plasmon thermal conductivity of Au and Ag thin films. Our theoretical predictions revealed a decrease of approximately 30\% in plasmon thermal conductivity when considering the size effect of the permittivity of thin metal films.  Furthermore, this causes the film thickness at which maximum thermal conductivity occurs to increase by about 30\%. Taking these factors into account, we calculated the optimal thickness of Au and Ag films, along with Ti adhesive layers, on SiO$_2$ substrates to be approximately 20 nm. By fabricating a sample with the optimal thickness of Au and Ag films, we experimentally demonstrated that the plasmon thermal conductivity of Au and Ag films can be as high as about 20\% of their electron contribution. This research will broaden the thermal design applications of ballistic thermal transport by SPPs propagating along thin metal coatings in microelectronics.
\end{abstract}


\maketitle


\section{Introduction}
For nanoscale thin films, the thermal conductivity of primary heat carriers such as phonons and electrons decreases as film thickness decreases due to their boundary scattering, which causes critical thermal issues for the performance of modern electronic devices \cite{qiu1993heat}. In response to this issue of the classical size effect, recent research efforts have shifted their focus towards exploiting ballistic thermal transport facilitated by surface electromagnetic waves (SEWs) \cite{chen2005surface, fu2006nanoscale, ordonez2013anomalous, ordonez2014thermal, lim2019thermal}. SEWs emerge from the coupling of infrared photons with either optical phonons (in polar dielectrics) or free electrons (in conductors), possessing propagation lengths that surpass those of the primary heat carriers in thin films by several orders of magnitude \cite{burke1986surface, yang1991long, joulain2005surface}. By harnessing SEWs as additional heat carriers, it becomes possible to compensate for the classical size effect observed in the thermal conductivity of thin films. Surface phonon polaritons (SPhPs), a specific class of SEWs generated through the coupling of infrared photons with the optical phonons in polar materials, have been theoretically proposed to enhance the thermal conductivity of a 40-nm-thick SiO$_2$ membrane by over 100\% compared to its bulk value \cite{chen2005surface}.

Previous investigations have extensively explored the phenomenon of ballistic thermal transport facilitated by SPhPs. Through analytical methods, \citet{ordonez2013anomalous} demonstrated that the thermal conductivity of suspended polar membranes can be significantly enhanced, reaching up to 1.8 times its bulk counterpart, by carefully selecting substrate permittivity, film thickness, and temperature. Subsequent experimental studies corroborated the thermal conductivity of suspended membranes in relation to varying film thickness \cite{tranchant2019two}, temperature \cite{wu2020enhanced}, and film width \cite{wu2022observation}. However, practical applications face challenges in realizing the full potential of SPhPs; that is, fabricating nanoscale-thick membranes larger than their propagation length poses difficulties in maintaining the robustness of this mode. Additionally, although SPhPs can propagate in supported multilayer polar films (i.e., an asymmetric medium), they exhibit well-confined behavior only when the thin film exceeds a characteristic thickness determined by the substrate and superstrate permittivity \cite{ordonez2014thermal, lim2019thermal, tachikawa2022plane}. With such a thick film, the heat flux carried by SPhPs becomes insignificant.

In contrast, surface plasmon polaritons (SPPs) represent an alternative form of SEWs and can also serve as heat carriers. SPPs arise from the interaction of photons with free electrons in metals. Unlike polar dielectric films, long-range SPPs can propagate over centimeters on `supported' metallic nanofilms in the mid-infrared regime \cite{burke1986surface, yang1991long}. Due to the high plasma frequency and large negative value of the real part of the metal permittivity, the frequency range of metal films that support SPPs (i.e., $\omega<\omega_p/\sqrt{2}$) is broader compared to that of polar dielectric films \cite{ordonez2023plasmon}. The thermally excited SPPs in the broad spectrum have been employed in cross-plane heat transfer, enabling control of near-field thermal radiation \cite{fu2006nanoscale, lim2018tailoring}. However, while the broad frequency range of SPPs holds potential for enhancing heat conduction in metal films, previous studies primarily focused on the propagation and thermal effects of SPPs at specific frequencies, overlooking the broader spectral characteristics \cite{lamprecht2001surface, garcia2005surfaces}. Recently, \citet{kim2023boosting} demonstrated the enhanced in-plane thermal conductivity of a Ti film by exploiting the long-range SPPs. It was shown that the plasmon thermal conductivity of a 100-nm-thick Ti film can contribute to additional energy transport by 25\% of its intrinsic thermal conduction by electrons and phonons. Despite the potential to achieve full propagation of SPPs on supported films, Ti films still face limitations in addressing thermal issues in microelectronics due to their low intrinsic thermal conductivity.

Alternatively, \citet{ordonez2023plasmon} theoretically investigated the plasmon thermal conductivity of a gold nanofilm on a silicon substrate, considering the effects of film thickness and temperature. They revealed that the plasmon thermal conductivity reaches about 25\% of its electron counterpart. However, experimental verification of the plasmon thermal conductivity in noble metal films, such as gold and silver, necessitates additional considerations. Firstly, the use of SiO$_2$ as a substrate is recommended to accurately measure the in-plane thermal conductivity of nanoscale metal films, as the low phonon thermal conductivity of SiO$_2$ enhances the measurement sensitivity \cite{kim2023boosting}. Secondly, the influence of an additional adhesive layer (e.g., Ti) on the dispersion of SPPs should be taken into account, particularly because noble metals exhibit poor adhesion to a glass substrate. Lastly, it is crucial to analyze the size effect of permittivity and its impact on the plasmon thermal conductivity, given that the film thickness at the maximum plasmon thermal conductivity for Au ($\sim 10$ nm \cite{ordonez2023plasmon}) is certainly shorter than the mean free path of electrons ($\sim 50$ nm).

In this study, we measure the plasmon thermal conductivity of nanoscale Au and Ag films deposited on a SiO$_2$ substrate with a Ti adhesive layer. The dispersion relation of SPPs propagating along the Au and Ag thin films is determined through numerical analysis, taking into account the size-dependent behavior of metal permittivity. Additionally, we examine the influence of the Ti adhesive layer on the SPP thermal conductivity as well as the thickness of Au and Ag films that lead to the maximum plasmon thermal conductivity. We achieve this by deriving the spatial distribution of SPPs along the films that incorporate the Ti adhesive layer. Through the fabrication of a sample with optimized Au and Ag film thicknesses, we conduct experimental investigations to explore the dependence of plasmon thermal conductivity on film thickness and radius.


\section{Theoretical model}

\begin{figure}[t!]%
\centering
\includegraphics[width=0.5\textwidth]{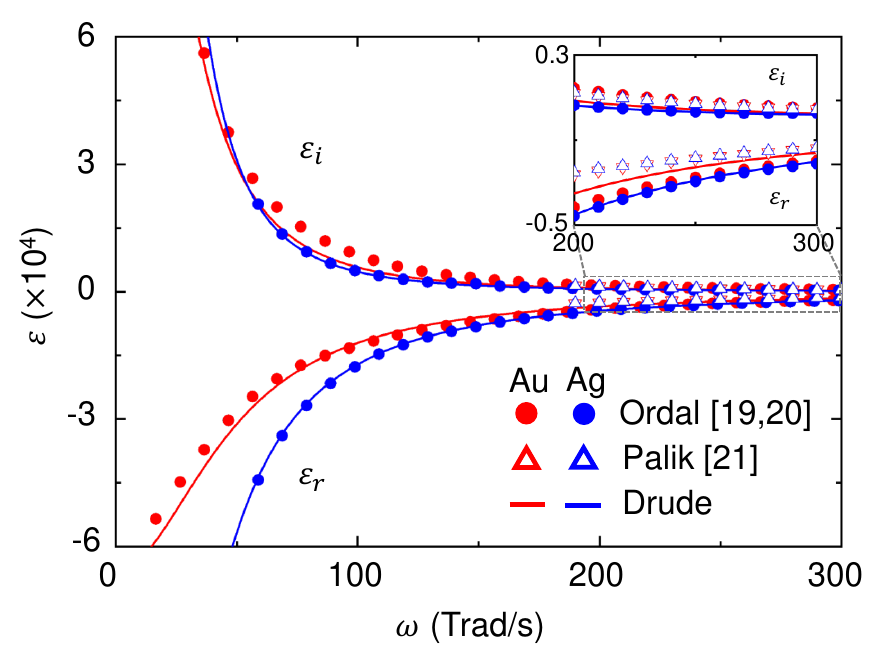}
\caption{Fitting of the Drude model to the existing experimental data of Au and Ag (bulk).}\label{fig:permittivity_fitting}
\end{figure}

We first found out how thick a film needs to be for the plasmon thermal conductivity to be at its highest. This made it easier to do experiments on the plasmon thermal conductivity of Au and Ag thin films. The calculation of plasmon thermal conductivity as a function of film thickness involved numerically solving the dispersion relation of SPPs propagating along the films. The dispersion relation for SPPs supported by a single metal film on a glass substrate (i.e.. no Ti adhesive layer) can be expressed \cite{ordonez2014thermal}:
\begin{equation} \label{Eq:3layer}
	 \tanh (p_md) = -\frac{p_{m}\varepsilon_{m}(p_a\varepsilon_s+p_s\varepsilon_a)}{p_{m}^2\varepsilon_a\varepsilon_s+p_ap_s\varepsilon_{m}^2},
\end{equation}
where, the subscripts `m', `a', and `s' represent the metal, air, and substrate, respectively, while $p_i$ denotes the cross-plane wave vector of the respective medium. The permittivity of each medium, denoted as $\varepsilon_i$, is used to numerically obtain the in-plane wave vector $\beta$ that satisfies $p_i^2=\beta^2-\varepsilon_i k_0^2$ for each medium.

The permittivity of `bulk' Au and Ag over a range of frequencies was predicted theoretically using the Drude model, i.e., 
\begin{equation} \label{Eq:drude}
	 \varepsilon_{\text{bulk}}=1-\frac{\omega_p^2}{\omega^2+i\omega \Gamma}.
\end{equation}
Here, the Drude model parameters, namely the damping coefficient $\Gamma$ and plasma frequency $\omega_p$, were fitted to the tabulated data \cite{ordal1985optical, ordal1987optical, palik1998handbook} using the least-squares method. The data from \citet{palik1998handbook} cover a spectral range exceeding 190 THz, so the Drude parameters were determined by fitting the data from \citet{ordal1985optical, ordal1987optical} below 190 THz. As a result, the Drude parameters for Au were determined as $\Gamma=252$ cm$^{-1}$ and $\omega_p=64660$ cm$^{-1}$, while for Ag, they were $\Gamma=145$ cm$^{-1}$ and $\omega_p=72071$ cm$^{-1}$. The inset of Fig.\ \ref{fig:permittivity_fitting} demonstrates that the theoretical prediction of the real part of the permittivity based on the Drude model falls between the measurement data from  \citet{palik1998handbook} and \citet{ordal1985optical, ordal1987optical}, and the imaginary part of the permittivity agrees well with both data sets. In addition, the permittivity of the SiO$_2$ substrate is determined using the tabulated data \cite{palik1998handbook}.

\begin{figure}[b!]%
\centering
\includegraphics[width=0.5\textwidth]{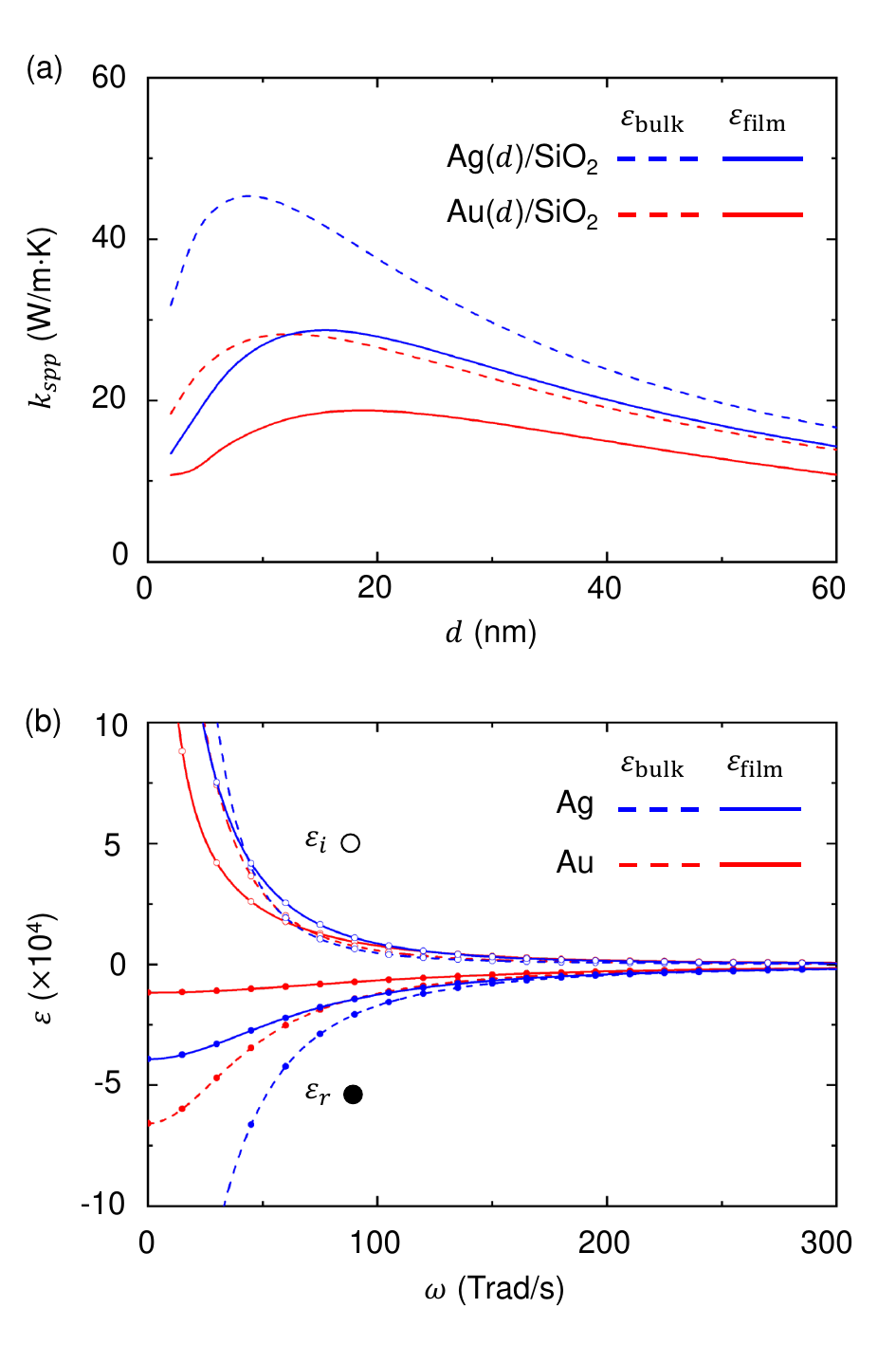}
\caption{(a) Calculated plasmon thermal conductivity of Au and Ag films directly deposited on a SiO$_2$ substrate with respect to film thickness when $L=50$ mm. (b) Real ($\varepsilon_r$) and imaginary ($\varepsilon_i$) part of permittivity when $d=16$ nm. Dashed lines are calculated based on the bulk permittivity of Au and Ag in Eq.\ \eqref{Eq:drude}, and solid lines are calculated based on the thin-film permittivity of Au and Ag with consideration of the size effect in Eq.\ \eqref{Eq:knudsen}. }\label{fig:size_effect}
\end{figure}

The plasmon thermal conductivity of a metal film on a silicon dioxide (SiO$_2$) substrate can be calculated by \cite{chen2005surface}
\begin{equation} \label{Eq:conductivity}
	 k_{\parallel,\text{spp}}=\frac{1}{4\pi d}\int^{\infty}_0 k_\omega d\omega=\frac{1}{4\pi d}\int^{\infty}_{0}\hbar\omega \Lambda_{\text{eff}}\beta_R\frac{df_0}{dT}d\omega,
\end{equation}
where $\hbar$ is the Planck constant divided by $2\pi$, $d$ is the thickness of the metal film, $\omega$ is the angular frequency, and $f_0$ corresponds to the Bose-Einstein distribution function. The real part of the in-plane wave vector is denoted as $\beta_R$, and the propagation length of SPPs is defined as $\Lambda=1/(2\beta_I)$ with $\beta_I$ being the imaginary part of the in-plane wave vector. The in-plane wave vector $\beta$ can be obtained by solving Eq.\ \eqref{Eq:3layer}. Finally, the effective propagation length ($\Lambda_{\text{eff}}$) for a metal film with a finite width $L$ is predicted by the Boltzmann transport equation \cite{guo2021quantum}.

Using the bulk permittivity in Eq.\ \eqref{Eq:drude}, the dashed lines in Fig.\ \ref{fig:size_effect}a show how the plasmon thermal conductivity of Au and Ag films directly deposited on a SiO$_2$ substrate (no Ti adhesive layer) changes with their thickness when  $L=50$ mm. At first, a quick comparison with the plasmon thermal conductivity of a Ti film on a SiO$_2$ substrate \cite{kim2023boosting} shows that the maximum plasmon thermal conductivity of Au and Ag films is more than 10 times higher than that of the Ti film. Furthermore, the corresponding thickness of Au and Ag films is approximately 7 times smaller than that of the Ti film. Remarkably, both Au and Ag films exhibit their maximum plasmon thermal conductivity at a thickness of approximately 10 nm. Considering that the plasmon thermal conductivity in thin metal films increases as the SPPs at both interfaces of the metal film are decoupled \cite{kim2023boosting}, these findings suggest that SPPs are less confined in Au and Ag films compared to Ti films. This indicates a shorter penetration depth into the metal and a deeper propagation into lossless air for Au and Ag films, implying a longer propagation length for SPPs and thus a higher plasmon thermal conductivity compared to Ti film.

According to \citet{gall2016electron}, the mean free path of electrons in Au is approximately 37.7 nm, while in Ag, it is approximately 53.5 nm. This implies that the boundary scattering of electrons within Au and Ag films, which is thickness-dependent, can result in variations in their permittivity. Thus, it becomes essential to account for these size-dependent permittivity when calculating the plasmon thermal conductivity. In the Drude model, the damping coefficient $\Gamma$ is inversely proportional to the relaxation time $\tau$, which represents the average time electrons travel between collisions \cite{zhang2007nano}. Specifically, the electron mean free path ($l$) can be expressed as $l = v_F \tau$, where $v_F$ denotes the Fermi velocity. In bulk metals, the electron mean free path is determined by the relaxation times associated with electron-electron, electron-phonon, and electron-defect scatterings. However, in noble metals like gold (Au) and silver (Ag), electron-electron scattering plays a dominant role \cite{tong2019comprehensive}. When considering thin films, electron-boundary scattering becomes significant, and the electron mean free path in a thin film ($l_{\text{film}}$) can be effectively described by \cite{zhang2007nano}
\begin{equation} \label{Eq:knudsen}
	 \frac{l_{\text{film}}}{l_{\text{bulk}}}=\left(1+\frac{\text{Kn}}{\ln(\text{Kn})+1}\right)^{-1},
\end{equation}
where $l_{\text{bulk}}$ is the \textit{bulk mean free path} of electrons and $\text{Kn}=l_{\text{bulk}}/d$ represents the Knudsen number with $d$ being the film thickness. Using Eq.\ \eqref{Eq:knudsen}, we can obtain the modified damping coefficient of the Drude model for a thin metal film by $\Gamma_{\text{bulk}}/\Gamma_{\text{film}}=l_{\text{film}}/l_{\text{bulk}}$. 

Solid lines in Fig.\ \ref{fig:size_effect}b show the predicted permittivity $\varepsilon_{\text{film}}$ of a 16-nm-thick Au and Ag film by considering the size effect based on $\Gamma_{\text{film}}$. The real part of the thin-film permittivity exhibits a significant reduction in the frequency range below 100 THz. Due to the reduced permittivity of Au and Ag films, the maximum value of the plasmon thermal conductivity when using $\varepsilon_{\text{film}}$ is approximately 30\% lower compared to that calculated with $\varepsilon_{\text{bulk}}$, as shown in Fig.\ \ref{fig:size_effect}a. Additionally, the film thickness at which the maximum plasmon thermal conductivity occurs for Au and Ag films, based on $\varepsilon_{\text{film}}$, is approximately 30\% larger than that based on $\varepsilon_{\text{bulk}}$. Figure \ref{fig:size_effect} provides clear evidence that the effects of electron-boundary scattering must be considered when modeling the permittivity for the purpose of analyzing the plasmon thermal conductivity of a noble metal film.

\begin{figure}[t!]%
\centering
\includegraphics[width=0.5\textwidth]{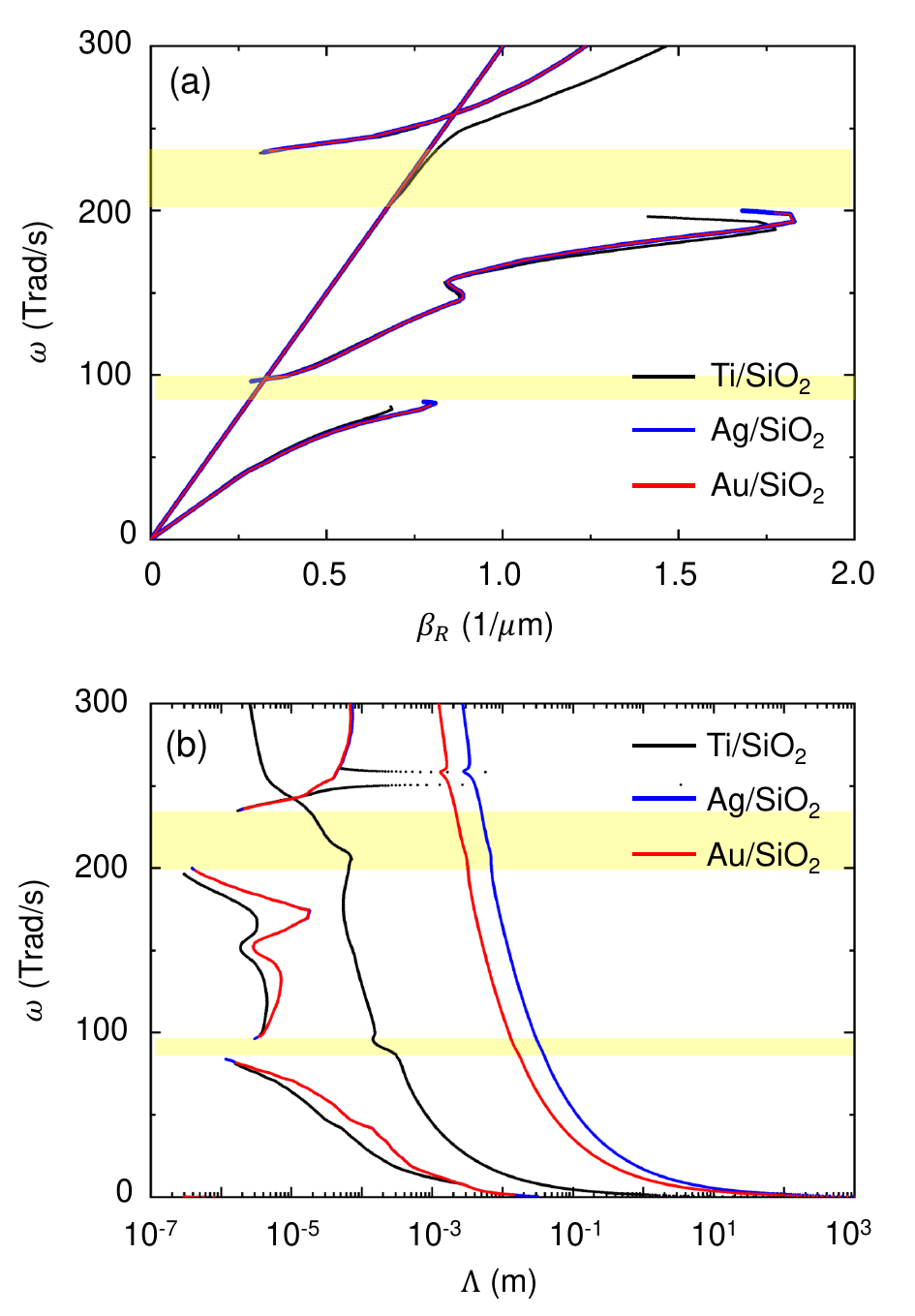}
\caption{(a) Real part of the in-plane wave vector of SPPs for 16-nm-thick Au, Ag, and Ti films directly deposited on a SiO$_2$ substrate. The shaded region indicates the frequency range where the real part of the dielectric function of the SiO$_2$ substrate becomes negative. (b) Propagation length of SPPs for 16-nm-thick Au, Ag, and Ti films directly deposited on a SiO$_2$ substrate.}\label{fig:beta}
\end{figure}

In order to examine the relationship between plasmon thermal conductivity and film thickness, which also affects the permittivity, Fig.\ \ref{fig:beta} illustrates the behavior of the real part of the in-plane wave vector $\beta_R$ and the propagation length $\Lambda_{\text{eff}}$ for Au, Ag, and Ti films directly deposited on a SiO$_2$ substrate (i.e., no adhesive layer) when $d=16$ nm and $L=50$ mm. The SPP dispersion curves are depicted within a frequency range from 0 to 300 THz, as the spectral plasmon thermal conductivity of the metal films is primarily concentrated below 300 THz \cite{kim2023boosting}. These dispersion curves originate from the light line of the surrounding mediums, namely $\omega/c=\varepsilon_{a,s}k_0^2$. Considering that thermal transport via SPPs in metal films is predominantly governed by the metal-air interface \cite{kim2023boosting}, the focus is primarily on the propagation of SPPs along the metal-air interface. Although it is not noticeable from Fig.\ \ref{fig:beta}a,  $\beta_R$ of Au and Ag films is closer to the light line of air than that of Ti film. Also, noble metals usually have low losses compared to refractory metals. Consequently, the propagation length of SPPs along the metal-air interface is approximately two orders of magnitude longer for Au and Ag films compared to Ti film, as depicted in Fig.\ \ref{fig:beta}b. 

Because $\beta_R$ approaches the light line of air (i.e., $\beta_R \sim k_0$) for the SPPs propagating along the metal-air interface, the real part of the cross-plane wave vector can be approximated as $p_m\sim k_0\sqrt{1-\varepsilon_{\text{film}}}$. Given that the permittivity of the metal is significantly larger than that of air (i.e., $\varepsilon_{\text{film}} \gg 1 $), the penetration depth of SPPs into the metal is primarily determined by the real part of the metal permittivity (i.e., $\delta_m=1/[2\text{Re}(p_m)]\sim1/[2\text{Re}(k_0\sqrt{-\varepsilon_{\text{film}}})]$). Consequently, the penetration depth of SPPs into Ti film is longer compared to Au and Ag films, because the real part of the Ti permittivity is much smaller than the real part of the Au and Ag permittivity. Therefore, the film thickness at which the maximum plasmon thermal conductivity occurs is much smaller for Au and Ag films compared to Ti film, as the maximum occurs when SPPs at both interfaces are decoupled (i.e., $d\sim 2\delta_m$). Due to the above reasons, Au and Ag films directly deposited on a SiO$_2$ substrate achieve the maximum plasmon thermal conductivity with the smaller film thickness compared to Ti films, as depicted in Fig.\ \ref{fig:size_effect}a. 

The plasmon thermal conductivity of the Au and Ag films, as demonstrated in Fig.\ \ref{fig:size_effect}b, exhibits a diminished peak value and an increased film thickness at the point of maximum conductivity when $\varepsilon_{\text{film}}$ is employed. The reduction in plasmon thermal conductivity can be explained quantitatively by examining the dispersion curves depicted in Fig.\ \ref{fig:beta}a. These dispersion curves for $\beta_R$ closely coincide with the analytical solutions representing the dispersions of SPPs at the single interface between a metal and its surrounding medium \cite{kim2023boosting}. The expressions for these dispersions are simply given by
\begin{equation} 
\label{Eq:analytic_betaR}
\beta_{a,s}=\varepsilon_{a,s}^{1/2}k_0\left(\frac{\varepsilon_{\text{film}}}{\varepsilon_{\text{film}}+\varepsilon_{a,s}}\right)^{1/2},
\end{equation}
where $\beta_a$ and $\beta_s$ represent the in-plane wave vectors at the metal-air and metal-substrate interfaces, respectively. Assuming that $\beta_R$ is significantly greater than $\beta_I$, the imaginary part of the in-plane wave vectors at each interface can be expressed by \cite{burke1986surface}:
\begin{equation} 
\label{Eq:analytic_betaI}
\beta_{a,sI}=\frac{1}{2}\frac{\varepsilon_{a,s}^2 \text{Im}(\varepsilon_{\text{film}})}{\beta_{a,sR}}\frac{k_0^2}{|\varepsilon_{\text{film}}+\varepsilon_{a,s}|^2},
\end{equation}
where $\text{Im}()$ takes the imaginary part. Because the plasmon thermal conductivity is proportional to $\beta_R/\beta_I$, as indicated in Eq.\ \eqref{Eq:conductivity}, and remains within the same order of magnitude across film thicknesses ranging from 0 nm to 60 nm (see Fig.\ \ref{fig:size_effect}a), it is reasonable to perform a quantitative analysis of the plasmon thermal conductivity using Eq.\ \eqref{Eq:analytic_betaI}. This analysis holds true regardless of whether the film falls under the category of thin or thick films, i.e., the coupled/decoupled regime.

As the film thickness decreases (resulting in an increase in $\Gamma_{\text{film}}$), the metal permittivity $\varepsilon_{\text{film}}$ decreases for both the real and imaginary parts. By assuming that $\omega_p$ is much greater than $\Gamma_{\text{film}}$ and by analytically separating the real and imaginary components of the Drude model from Eq.\ \eqref{Eq:drude}, it can be observed that the real part of $\varepsilon_{\text{film}}$ is inversely proportional to $\Gamma_{\text{film}}^2$, while the imaginary part of $\varepsilon_{\text{film}}$ is inversely proportional to $\Gamma_{\text{film}}$. Hence, as the size effect leads to an increase in $\Gamma_{\text{film}}$, the propagation length of SPPs on the metal-air interface diminishes, thereby causing a decrease in the plasmon thermal conductivity. This reduction can be attributed to the amplified $\beta_{aI}$ term (as shown in Eq.\ \eqref{Eq:analytic_betaI}), primarily driven by the dominant increase in $\varepsilon_{\text{film}}$. Furthermore, the increase in film thickness at the point of maximum plasmon thermal conductivity can also be elucidated by considering the skin depth of SPPs within the metal film. As mentioned previously, the skin depth in the metal film, denoted as $\delta_m$, is inversely proportional to the real part of $\varepsilon_{\text{film}}$. Consequently, considering the size effect, which leads to a decrease in the absolute value of the real part of $\varepsilon_{\text{film}}$, results in a slight augmentation of $\delta_m$. A larger $\delta_m$ signifies that a greater film thickness is necessary for the decoupling of SPPs at the interface. Therefore, when accounting for the size effect of the metal permittivity in thin films, the film thickness at the point of maximum plasmon thermal conductivity increases.

\section{Results and discussion}

\begin{figure}[b!]%
\centering
\includegraphics[width=0.5\textwidth]{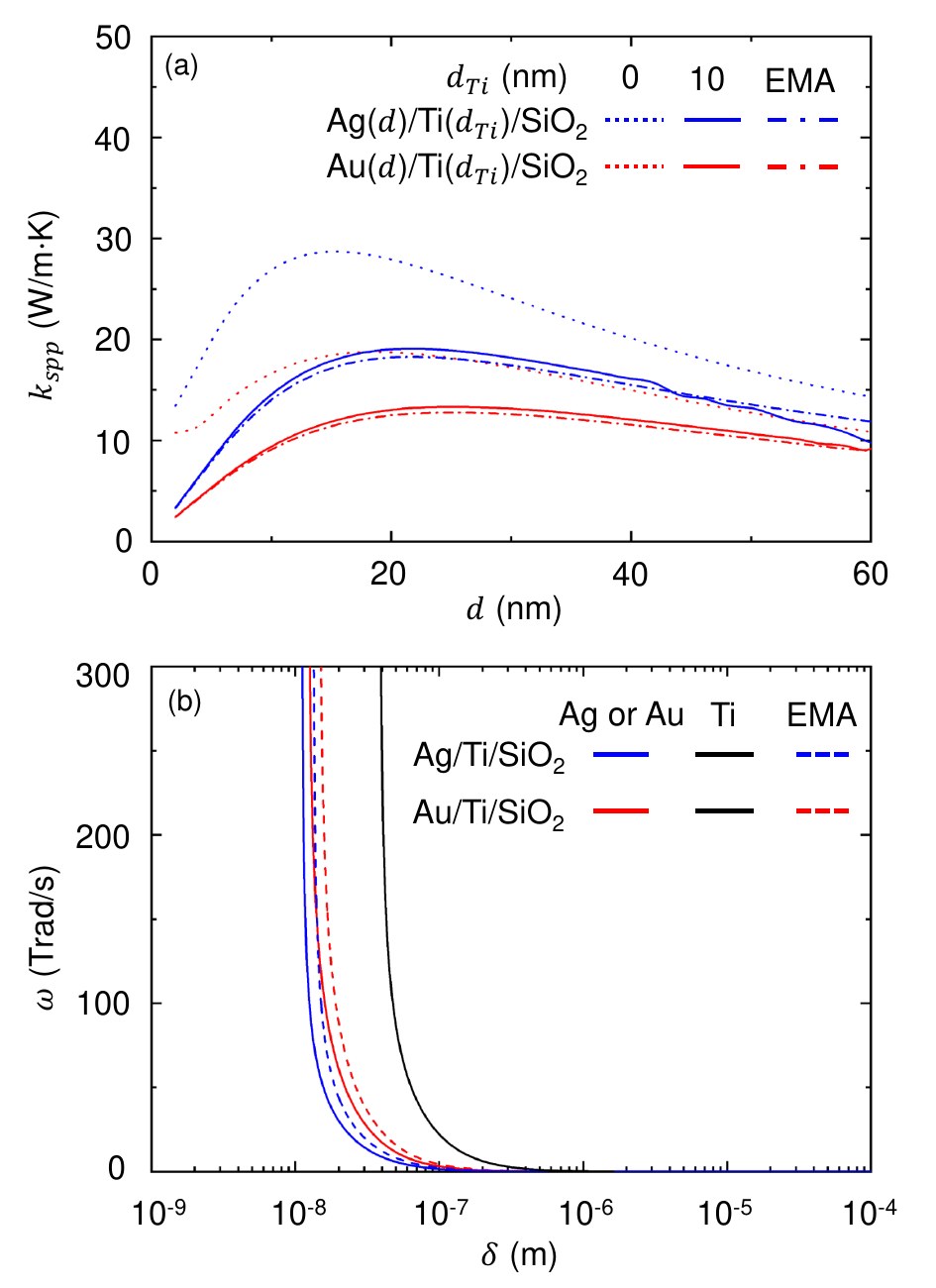}
\caption{(a) Plasmon thermal conductivity with respect to Au and Ag film thickness on a SiO$_2$ substrate with a 10-nm-thick Ti adhesive layer. EMA represents the calculated plasmon thermal conductivity of the effective medium consisting of Au or Ag film and Ti adhesive layer by using the effective medium approximation. (b) Skin depth of SPPs along 20-nm-thick Au and Ag films, and 10-nm-thick Ti adhesive layer.}\label{fig:Ti_effect}
\end{figure}

The relatively low thermal conductivity of SiO$_2$ enhances the measurement sensitivity of in-plane thermal conductivities in Au and Ag films by concentrating the heat transfer pathway on the metal films. However, due to the weak adhesion between the Au and Ag films and the SiO$_2$ substrate, an additional Ti adhesive layer must be deposited between the metal films and the substrate. It is necessary for the Ti adhesive layer to have a minimum thickness of 10 nm to ensure sufficient adhesive force. By solving the dispersion relation of SPPs with the presence of the Ti adhesive layer, we determined the film thickness at which the maximum plasmon thermal conductivity occurs for the Au and Ag films. The dispersion relation of SPPs in a 4-layer structure can be expressed as follows \cite{ordonez2014thermal}:
\begin{equation} \label{Eq:4-layer}
	 \frac{\text{tanh}(p_1d_1)+\alpha_{01}}{1+\alpha_{01}\text{tanh}(p_1d_1)}=-\alpha_{21}\frac{\text{tanh}(p_2d_2)+\alpha_{32}}{1+\alpha_{32}\text{tanh}(p_2d_2)},
\end{equation}
where the subscript $i$ denotes the $i^{\text{th}}$ medium (i.e., $i=1,2,3,4$), and $\alpha_{ij}=\varepsilon_i p_j/\varepsilon_j p_i$. By substituting the solved value of $\beta$ from Eq.\ \eqref{Eq:4-layer} into Eq.\ \eqref{Eq:conductivity}, the calculated plasmon thermal conductivity of Au and Ag films with a Ti adhesive layer is plotted in Fig.\ \ref{fig:Ti_effect}a. Upon introducing a 10-nm-thick Ti adhesive layer between the Au and Ag films and the substrate, the plasmon thermal conductivity decreases by approximately 25\%, and the film thickness at which the maximum plasmon conductivity occurs increases by around 30\%. 

The impact of the Ti adhesive layer on the plasmon thermal conductivity can be explained by considering the skin depth of SPPs within the metal ($\delta_m$). Figure \ref{fig:Ti_effect}b illustrates the skin depth of SPPs in the 10-nm-thick Ti adhesive layer when the film thickness of Au and Ag films is 20 nm. For frequencies above 200 THz (corresponding to the peak wavelength of the Planck distribution at 300 K), the skin depth of SPPs in the Ti adhesive layer exceeds 40 nm. Consequently, there is relatively little attenuation of SPPs within the Ti adhesive layer. In other words, the dispersion of SPPs along the metal films is expected to be negligibly affected by the Ti adhesive layer. Considering that the majority of thermal transport through SPPs occurs at the metal-air interface, the presence of the Ti adhesive layer primarily influences the overall film thickness used in the calculation of plasmon thermal conductivity in Eq.\ \eqref{Eq:conductivity}. As a result, the plasmon thermal conductivity decreases since it is inversely proportional to the overall film thickness. Conversely, the film thickness at which the maximum plasmon thermal conductivity occurs increases with the Ti adhesive layer due to the augmented effective skin depth within the metal films. To estimate the effective skin depth of SPPs in Au and Ag films with a Ti adhesive layer, we employed the effective medium approximation (EMA) for thin films and evaluated the effective permittivity as proposed by \citet{aspnes1982optical}
\begin{equation} \label{Eq:EMA}
	 \varepsilon_{\text{eff}}=f_{\text{Au,Ag}}\varepsilon_{\text{Au,Ag}}+f_{\text{Ti}}\varepsilon_{\text{Ti}},
\end{equation}
where $f$ represents the volume fraction (i.e., film thickness ratio) of the corresponding medium. Note that we considered the electron-boundary scattering effects for both Au (or Ag) and Ti layers when estimating their permittivity by using Eq.\ \eqref{Eq:knudsen}. Since Au and Ag have greater permittivity than Ti, the effective permittivity $\varepsilon_{\text{eff}}$ decreases proportionally to the film thickness ratio of Au and Ag films. As a consequence, the effective skin depth of Au and Ag films with a Ti adhesive layer is approximately 25\% larger (around 15 nm) compared to that of single Au and Ag films (around 12 nm). Furthermore, the effective skin depth exhibits a similar magnitude to the radiation penetration depth at the peak wavelength of the Planck distribution at 300 K ($\sim$10 $\mu$m), which corresponds to about 15 nm.

\begin{table}[b!]
\centering
\caption{\color{black} Measured values of $k_{\parallel,e}$ from the 4-probe method.}
\label{tab1}
{\begin{tabular}{c|c|c|c|c|c|c} \hline \hline

  & \multicolumn{3}{c|}{Au ($d=22.2\pm0.4$ nm)}  & \multicolumn{3}{c}{Ag ($d=20.1\pm1.4$ nm)}   \\ \hline

radius     & \begin{tabular}[c]{@{}c@{}} $R_s$\\ (Ohm/sq)\end{tabular} & \begin{tabular}[c]{@{}c@{}}$\rho$\\ (Ohm$\cdot$m)\end{tabular} & \begin{tabular}[c]{@{}c@{}}$k_{\parallel,e}$ \\ (W/m$\cdot$K)\end{tabular} 
   & \begin{tabular}[c]{@{}c@{}}$R_s$\\\ (Ohm/sq)\end{tabular} & \begin{tabular}[c]{@{}c@{}}$\rho$\\ (Ohm$\cdot$m)\end{tabular} & \begin{tabular}[c]{@{}c@{}}$k_{\parallel,e}$ \\ (W/m$\cdot$K)\end{tabular}  
 \\ \hline \hline

\centering \textbf{50 mm}  & 3.78  & 8.41$\times$10$^{-8}$ & 81.59 &  2.94 & 5.91$\times$10$^{-8}$ & 110.42           \\ \hline

\centering \textbf{20 mm}   & 4.27  & 9.47$\times$10$^{-8}$ & 72.38 &  3.02 & 6.07$\times$10$^{-8}$ & 107.57             \\ \hline

\centering \textbf{10 mm}   & 3.94  & 8.76$\times$10$^{-8}$ & 78.27 & 2.98 & 5.99$\times$10$^{-8}$ & 108.90            \\ \hline

\centering \textbf{Avg.}   &   &  & \textbf{77.4$\pm$4.7} &  &  & \textbf{108.7$\pm$1.4}   \\ \hline \hline

\end{tabular}
}
\end{table}

From our calculations of the plasmon thermal conductivity of Au and Ag films with a Ti adhesive layer, we found that 20 nm is the best thickness for both Au and Ag films. After that, we made the optimized sample by using e-beam evaporation to deposit Au and Ag films 20 nm thick and a Ti adhesive layer 10 nm thick on a SiO$_2$ substrate. The deposited films were patterned in circular shapes with various radii ranging from 500 $\mu$m to 50 mm, following a similar approach as described in our previous work \cite{kim2023boosting}. We measured the roughness of the Au and Ag film surfaces and found that it was less than 0.5 nm (see Section 1 of the Supplementary Material for more information). This showed that the conditions for long-range SPPs could be met. Furthermore, it is known that an Au film with a thickness of 20 nm follows the optical properties predicted by the simple mean free path theory (i.e., $\Gamma_{\text{film}}$ with size effect) \cite{stenzel2019spectrophotometric}. The thicknesses of the Au and Ag films were measured using a stylus profiler (Alpha-Step 500, KLA TENCOR CORP.). The measured thicknesses were found to be $22.2 \pm 0.4$ nm for Au films and $20.1 \pm 1.4$ nm for Ag films. To determine the electron thermal conductivity $k_{\parallel,e}$ of the Au and Ag films, we utilized sheet resistances measured via the 4-probe method (4200-SCS, Keithley). Both Au and Ag films follow the Wiedemann-Franz law, which states that thermal conductivity and electrical conductivity are proportional, given by the equation:
\begin{equation} \label{Eq:WFlaw}
	 \frac{k_{\parallel,e}}{\sigma}=L_mT,
\end{equation}
where $L_m$ is the Lorentz number for the respective metals. The Lorentz number is known to be 2.51$\times10^{-8}$ W$\cdot\Omega$/K$^2$ for Au and 2.39$\times10^{-8}$ W$\cdot\Omega$/K$^2$ for Ag \cite{tong2019comprehensive}. Here, $\sigma=1/\rho=1/(R_sd)$ represents the electrical conductivity, where $R_s$ is the sheet resistance and $d$ is the film thickness. Using the measured data of sheet resistance and film thickness, we obtained the electron thermal conductivity of the deposited Au and Ag films as $77.4 \pm 4.7$ W/m$\cdot$K and $109.0 \pm 1.4$ W/m$\cdot$K, respectively. Additionally, since the penetration depth of SPPs into air is greater than 100 $\mu$m, the presence of the native oxide layer on the Au and Ag films does not significantly affect the propagation of long-range SPPs \cite{ditlbacher2002two}.


\begin{figure}[ht!]%
\centering
\includegraphics[width=0.5\textwidth]{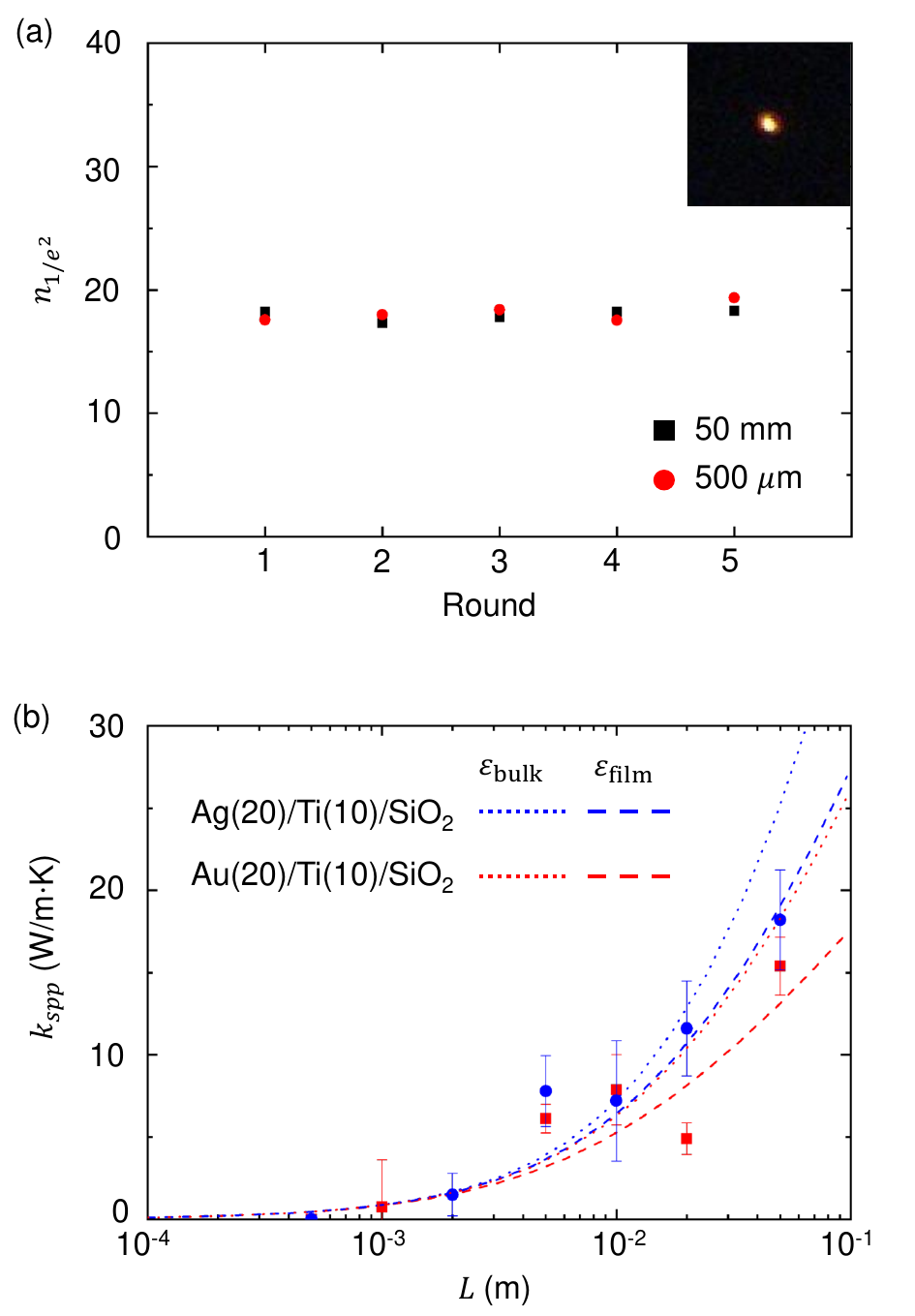}
\caption{(a) Number of pixel with intensity over $1/e^2$ (Gaussian diameter) from the beam image of $L=500$ $\mu$m and $L=50$ mm sample during five measurements. (b) Measured $k_{spp}$ with respect to film width for Au/Ti/SiO$_2$ and Ag/Ti/SiO$_2$ samples.}\label{fig:exp}
\end{figure}

The plasmon thermal conductivity was measured using the steady-state thermoreflectance method (SSTR) \cite{braun2019steady,kim2023boosting}. For the measurements, a film with a radius of 500 $\mu$m served as the calibration sample. By comparing the steady-state temperature rise induced by the pump laser at a heated spot between the measurement sample and the calibration sample, it is possible to extract the change in the in-plane thermal conductivity of the measurement sample caused by the plasmon thermal conductivity. To determine the plasmon thermal conductivity of Au and Ag films, a 2-D heat diffusion model was employed, incorporating the measured data of temperature rise, beam size, film thickness, and bulk thermal conductivity of the films and substrate \cite{braun2017upper,kim2023boosting}. In the case of Au and Ag films, the phonon thermal conductivity \cite{tong2019comprehensive} and the measurement sensitivity of the cross-plane thermal conductivity \cite{kim2023boosting} can be considered negligible, thus enabling the utilization of the film thermal conductivity values from Table 1 within the 2-D heat diffusion model. Also, due to the low sensitivity of boundary conductance $G$ in solving the 2-D heat diffusion model, its value is assumed to be 200 MW/m$^2$$\cdot$K \cite{braun2019steady, kim2023boosting}. The beam sizes of the probe and pump lasers were measured to be 4.8 $\mu$m and 4.7 $\mu$m, respectively, using the knife-edge method \cite{kim2023boosting}.

Maintaining consistency in the beam size during the measurement processes is essential, because it determines the heat flux from the heated spot. Hence, we continuously monitored the beam size using a CCD camera during five separate measurements of the largest sample ($L=50$ mm) and the calibration sample ($L=500$ $\mu$m) of Au, as illustrated in  Fig.\ \ref{fig:exp}a. By fitting the pixel intensity to a Gaussian distribution, we obtained the relative values of the Gaussian diameter of the lasers in terms of the number of pixels, denoted as $n_{1/e^2}$. The results indicated that $n_{1/e^2}$ was estimated to be $17.97 \pm 0.41$ for the largest sample and $18.18 \pm 0.75$ for the calibration sample. Although the beam sizes of the largest sample were slightly smaller than those of the calibration sample by less than 1\%, this difference is expected to have a negligible impact on the temperature rise of the calibration sample within the 2-D heat diffusion model. In fact, the measurement data presented in Fig.\ \ref{fig:exp}b reveal that the temperature rise of the heated spot in the measurement sample is lower compared to that of the calibration sample. This observation suggests that the change in beam size does not affect the observation of the plasmon thermal conductivity.

Figure  \ref{fig:exp}b shows the measured plasmon thermal conductivity ($k_{spp}$) of two samples: a 20 nm thick Au film and a 20 nm thick Ag film, both with a 10 nm thick Ti adhesive layer on a SiO$_2$ substrate. Both Au and Ag films exhibit a substantial dependence on the film radius ($L$). Specifically, the sample with a 50 mm radius for both Au and Ag films demonstrates a plasmon thermal conductivity of approximately 20\% relative to the electron thermal conductivity obtained from Table 1. For comparison purposes, theoretical predictions based on the kinetic theory, employing $\varepsilon_{\text{bulk}}$ and $\varepsilon_{\text{film}}$, are also depicted in Fig. \ref{fig:exp}b. Notably, the measured data closely aligns with the theoretical prediction based on the size effect of the metal permittivity. Due to the size effect of the metal permittivity, the theoretical prediction of $k_{spp}$ at $L=50$ mm exhibits a decrease of approximately 30\% compared to that obtained using $\varepsilon_{\text{bulk}}$. Furthermore, as illustrated in Fig. \ref{fig:Ti_effect}a, the calculated plasmon thermal conductivity values for 20-nm-thick Au and Ag films without a Ti adhesive layer amount to almost 30\% of their corresponding electron thermal conductivity. The boundary scattering of electrons reduces not only the plasmon thermal conductivity but also the intrinsic thermal conductivity. Thus, the ratio between plasmon and electron thermal conductivities remains relatively constant with respect to the size effect of metallic thin films. Moreover, this corresponding ratio closely resembles that found in the Ti film examined in a previous study by \citet{kim2023boosting}. This observation suggests a close relationship between plasmon and electron thermal conductivities in metallic films, possibly attributed to their mutual dependence on the electron number density, represented as $n_e$.

\section{Conclusion}
The SSTR measurements of supported Au and Ag films with a Ti adhesive layer on a SiO$_2$ substrate provide insights into the thickness and radius dependence of the plasmon thermal conductivity. Through the utilization of the Boltzmann transport equation based on the metal permittivity incorporating the size effect, we determined the optimal thickness of approximately 20 nm for both Au and Ag films to achieve the maximum plasmon thermal conductivity. Importantly, when considering the size effect of the permittivity, the theoretical prediction of the plasmon thermal conductivity experiences a reduction of approximately 30\%. For the Au and Ag films with $L=50$ mm, the plasmon thermal conductivity ($k_{spp}$) exhibits an increase of approximately 20\% compared to its electron counterpart. These experimental findings also provide validation for the mean free path theory of the permittivity in the supported structure of Au and Ag thin films. Overall, this study shows that plasmon thermal conductivity can be used in more real-world situations by using noble metal nanoscale-supported films.

\begin{acknowledgments}
This research is supported by the Basic Science Research Program (NRF-2019R1A2C2003605 and NRF-2020R1A4A4078930) through the National Research Foundation of Korea (NRF) funded by Ministry of Science and ICT.

\section*{Supplemental Material}
See Supplemental Material at [URL will be inserted by publisher] for (1) Surface roughness of Au and Ag thin films. All files related to a published paper are stored as a single deposit and assigned a Supplemental Material URL. This URL appears in the article's reference list.

\end{acknowledgments}


\bibliography{Kim_aps.bib}

%

\end{document}


\setstretch{1.5}
\section*{Supplemental Material:\\
Plasmon thermal conductivity of thin Au and Ag films}

\textbf{Authors}: Dong-min Kim$^{1,2}$, Jeongmin Nam$^{1,2}$, and Bong Jae Lee$^{1,2*}$\\

\textit{1. Department of Mechanical Engineering, Korea Advanced Institute of Science and Technology, Daejeon 34141, South Korea\\
2. Center for Extreme Thermal Physics and Manufacturing, Korea Advanced Institute of Science and Technology, Daejeon 34141, South Korea\\
}\\
*Corresponding authors\\
*e-mail: bongjae.lee@kaist.ac.kr (Bong Jae Lee)\\

\bigskip
\bigskip
\bigskip
\underline{\textbf{Table of Contents}}\\
1. Surface roughness of Au and Ag thin films\\

\setstretch{1.5}

\clearpage
\section*{1. Surface roughness of Au and Ag thin films}

\begin{figure}[!h]
\centering\includegraphics[width=17cm]{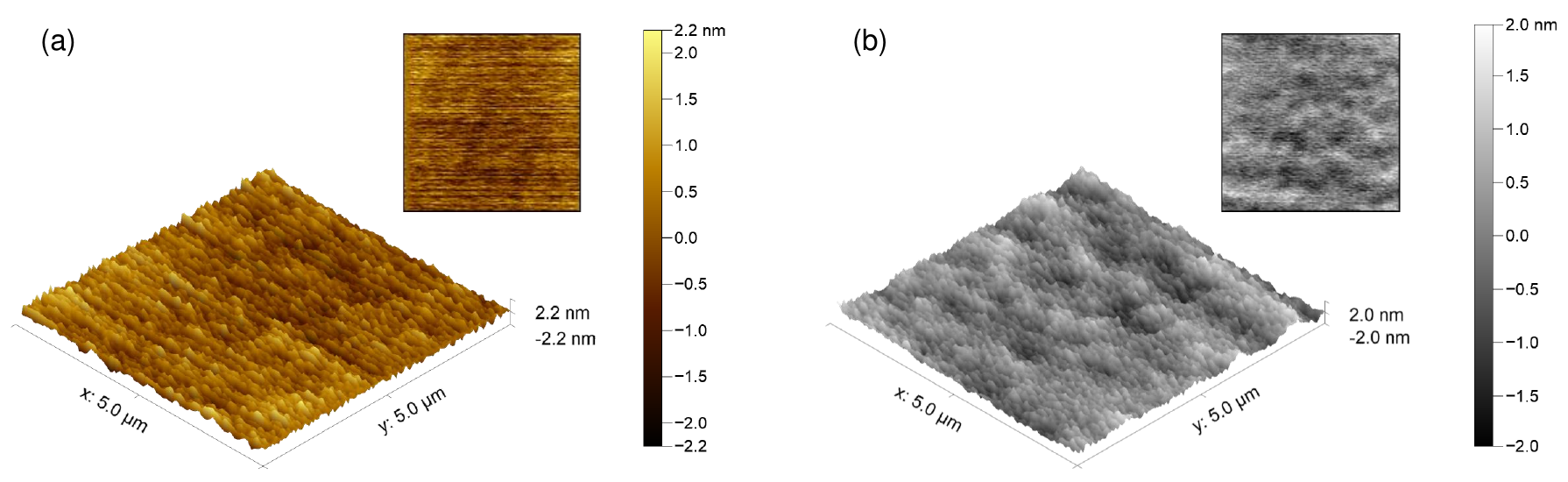}
\caption{\label{FigS1}Figure S1: (a) 5 $\mu$m $\times$ 5 $\mu$m AFM image of (a) Au surface and (b) Ag surface.}
\end{figure}

The surface roughness of Au and Ag thin films deposited on SiO$_2$ substrates was characterized. 5 $\mu$m $\times$ 5 $\mu$m surface images of the Au and Ag films were obtained with a conventional atomic force microscope (NTEGRA Aura, NT-MDT) and a Si probe (RESP10, Bruker). AFM images of the Au and Ag films are shown in Fig. S\ref{FigS1}. The difference of maximum and minimum values of the surface height differ by 4.4 nm for Au and 3.95 nm for Ag. The standard deviation of the surface height was measured to be 0.6 nm for Au and 0.5 nm for Ag. Roughness of the surface is commonly defined by arithmetical mean deviation as \cite{etxeberria2015useful}:
\begin{equation} \label{Eq:roughness}
	S_a=\frac{1}{A}\iint_{A} |Z(x,y)|dxdy,
\end{equation}
where $A$ is the area of the image, and $Z(x,y)$ is the mean deviation of film height at point ($x,y$). The calculated roughness of the films derived from Eq. \eqref{Eq:roughness} is 0.5 nm for Au and 0.4 nm for Ag, which are negligible compared to the SPP wavelength. Thus, the propagation of SPPs should not be attenuated by defects at the film interface.\\

\newpage

\bibliographystyle{naturemag}
\bibliography{Kim_supplemental_material}